\documentclass[english,aip,apl,amsmath,amssymb,reprint]{revtex4-1}
\usepackage{graphicx}
\begin{document}

\title{99.992\%   $^{28}$Si CVD-grown epilayer on 300 mm substrates for large scale integration of silicon spin qubits}
\author{V.~Mazzocchi}
\affiliation{CEA, LETI, Minatec Campus, F38054 Grenoble, France}
%\ead{vincent.mazzocchi@cea.fr}
\author{P.~G.~Sennikov}
\affiliation{G. G. Devyatykh IChHPS RAS , 603950 Nizhny Novgorod, Russian Federation}
%\ead{pgsen@rambler.ru}
\author{A.~D.~Bulanov}
\affiliation{G. G. Devyatykh IChHPS RAS , 603950 Nizhny Novgorod, Russian Federation}
\author{M.~F.~Churbanov}
\affiliation{G. G. Devyatykh IChHPS RAS , 603950 Nizhny Novgorod, Russian Federation}
\author{B.~Bertrand}
\affiliation{CEA, LETI, Minatec Campus, F38054 Grenoble, France}
\author{L.~Hutin}
\affiliation{CEA, LETI, Minatec Campus, F38054 Grenoble, France}
\author{J.~P.~Barnes}
\affiliation{CEA, LETI, Minatec Campus, F38054 Grenoble, France}
\author{M.~N.~Drozdov}
\affiliation{IPM RAS , 603950 Nizhny Novgorod, Russian Federation}
\author{J.~M.~Hartmann}
\affiliation{CEA, LETI, Minatec Campus, F38054 Grenoble, France}
\author{M.~Sanquer}
\affiliation{Univ. Grenoble Alpes, CEA, INAC-Pheliqs, 38000 Grenoble , France}
%\ead{marc.sanquer@cea.fr}
%\author[MINATEC]{V.~Mazzocchi\corref{cor2}\fnref{fn1,fn3}}
%\ead[url]{http://www.elsevier.com}
%\cortext[cor2]{Principal corresponding author}
%\fntext[fn1]{This is the specimen author footnote.}
%\fntext[fn2]{Another author footnote, but a little more
%longer.}
%\fntext[fn3]{Yet another author footnote. Indeed, you can have
%any number of author footnotes.}
%\address[MINATEC]{CEA, LETI, Minatec Campus, F38054 Grenoble, France}
%\address[IChHPS]{G. G. Devyatykh IChHPS RAS , 603950 Nizhny Novgorod, Russian Federation}
%\address[IPM]{IPM RAS , 603950 Nizhny Novgorod, Russian Federation}
%\address[cea]{Univ. Grenoble Alpes, CEA, INAC-Pheliqs, 38000 Grenoble , France}

\begin{abstract}
Silicon-based quantum bits with electron spins in quantum dots or nuclear spins on dopants are serious contenders in the race for quantum computation.
 Added to process integration maturity, the lack of nuclear spins in the most abundant $^{28}$silicon isotope host crystal for qubits is a major asset
  for this silicon quantum technology. We have grown $^{28}$silicon epitaxial layers (“epilayers”) with an isotopic purity greater than 99.992 \%
   on 300mm natural abundance silicon crystals. The quality of the mono-crystalline isotopically purified epilayer conforms to the same drastic quality requirements as the natural epilayers used in our pre-industrial CMOS facility. The isotopically purified substrates are now ready for the fabrication of silicon qubits using a state-of-the-art 300 mm Si CMOS-foundries equipment and processes
\end{abstract}


%\begin{keyword}
%D. quantum computing \sep A. enriched silicon \sep  B. CVD growth \sep B. 300 mm wafers
%\PACS 03.67.Lx (quantum computation) \sep 28.60.+s (isotopes separation and enrichment) \sep 85.40.-e (Microlectronics)
%A1. isotopes separation and enrichment  28.60.+s \sep
%A3. Chemical vapor deposition processes \sep
%B2. Semiconducting silicon \sep
%B3. Quantum devices, silicon spin quantum bits \sep
%B3. Microlectronics 85.40.-e
%\end{keyword}
\maketitle

 \section{Introduction}

Spin quantum bits in isotopically purified $^{28}$Si present remarquably longer (by two to four orders of magnitude) inhomogeneous  dephasing time as compared to their natural Si counterparts
(T$_2^*$=120$\mu$s for quantum bits based on MOS quantum dots\cite{Veldhorst14},  140$\mu$s/270$\mu$s for quantum bits based on donors\cite{Muhonen14}, 20$\mu$s for quantum bits based on $^{28}$Si-SiGe heterostructures \cite{Tarucha18}, for 99.92\% $^{28}$Si crystals).
Two--quantum bit--gates fabricated with isotopically purified 99.92\% $^{28}$Si exhibit also high fidelities \cite{Veldhorst15,Tarucha18}, much better than those based on natural Si\cite{Kawakami14,Takeda16}.

Recently qubits based on isotopically purified MOS structures based on 300 mm wafers covered with  a 100 nm thick layer $^{28}$Si purified at 99.92\%   have been reported \cite{Petit18}.  Long spin relaxation time T$_1$ ($\simeq 145 ms$) has been measured.

It is still unclear what are the mechanisms for limiting the dephasing time T$_2$ for silicon spin qubit  (residual $^{29}$Si nuclear spins, electrical noise mediated by spin-orbit or spin-valley coupling, etc.)  and if a higher level of isotopic purification is needed. 
In bulk $^{28}$Si ($^{29}$Si=0.005 \% ), very long  T$_2$ for electron spins on phosphorus donors   at a concentration of 1.2 10$^{14}$cm$^{-3}$  has been measured by electron spin resonance techniques:  T$_2$=600 ms, extrapolated to 10s by using a magnetic field gradient impeding flip-flop relaxation with residual nuclear $^{29}$Si spins \cite{Tyryshkin12}. The fundamental limit is T$_2$ $\simeq$ T$_1$=2000s at low temperature, indicating that progress can still be made if one reduces the $^{29}$Si content beyond 0.08\% in the silicon crystal used to date for  spin quantum bits \cite{Tyryshkin12}.
  
%The dependence of decoherence time on $^{29}$Si concentration has not  yet been tested for hole spin qubit in silicon

The general objective of the presented study is to  obtain $^{28}$Si crystalline layers with enrichment $\ge$99,992 \% and a very low level of other contaminants to fulfill the strict requirements of pre-industrial CMOS foundries.
A first CMOS qubit fabricated in a pre-industrial  foundry has been recently published by some of us \cite{Maurand16}.
 For instance the content of P and B dopants should be 10$^{12}$ cm$^{-3}$ or less, C and O contents 10$^{15}$ cm$^{-3}$ or less.
The high morphological quality $^{28}$Si  epilayers should be grown on 300 mm natural abundance Si substrates or silicon-on-insulator (SOI) substrates.
The supply chain should also provide enough isotopically purified materials to grow dozen’s of 300 mm diameter epilayers.

Previous achievements of $^{28}$Si enrichment are listed below for comparison.
The first sample of bulk $^{28}$Si (5 g) with  99.98 $\pm$ 0,02 \% enrichment and  a  4 $\times$ 10$^{16}$ cm$^{-3}$  phosphorous doping was produced by authors of ref. \cite{Feher58} for a study of spontaneous emission of radiation from an  electron spin system. The spin-lattice time T$_1$ and dephasing time  T$_2$ equal to 0.52 ms at T=1.4 K were measured for the first time in this sample  \cite{Gordon58}. 
The next attempt to manufacture bulk $^{28}$Si single crystal (300 g) by the reduction of $^{28}$SiO$_2$ with aluminum dates back to 1995 \cite{Becker95}. This attempt was not successful because of poor enrichment (99.02\%) and high concentration of boron and aluminum impurities. A method  based   on the reduction of $^{28}$SiF$_4$ with CaH$_2$ to $^{28}$SiH$_4$ was proposed in 2000\cite{Bulanov00}and used later e.g. in ref. \cite{Ruf00} and ref. \cite{Itoh03}.
 A similar approach has been used in \cite{Ager05}  without indication of the reducing chemical agent.  The enriched silane was purified by distillation and thermally decomposed to polycrystalline $^{28}$Si. The latter was used for growth of $^{28}$Si single crystal by float-zone or Czokhralski method. The enrichment of silicon depends on the enrichment of the starting $^{28}$SiF$_4$. Nowadays the best value is 99.99930 \% \cite{Abrosimov17} for  n-type $^{28}$Si single crystal with mass larger than 5 kg (concentration of $^{29}$Si 0.000658\%). For the  purest part of this crystal concentration of carbon and oxygen was less 10$^{15}$ cm$^{-3}$,  boron and phosphorus  less 10$^{13}$ cm$^{-3}$. In \cite{Sennikov12} the deposition of 525 $\mu$m layer of microcrystalline  $^{28}$Si with a 99.9986 \% enrichment on a $^{28}$Si (99.995 \%) substrate was achieved thanks to an  enriched $^{28}$SiF$_4$ in electron cyclotron resonance  discharge. Extremely high-enriched thin film (100-200 nm) of  $^{28}$Si 99.99983 \% with  $^{29}$Si concentration less  0.0001 \% was grown using hyperthermal  energy ion beam deposition system from SiH$_4$ of natural isotope abundance \cite{Dwyer14}. Films contained carbon and oxygen impurities at concentrations of the order of 3-4 \%. 

%For the goals of this study the quality of SiH$_4$, i.e its isotopic and chemical purity, is important. 
In studies  \cite{Takyu99,Ladd02,Shimizu07,Lo09} layers of $^{28}$Si were grown using $^{28}$SiH$_4$ produced in 1998-2005 by the Isonics Corporation, USA, in cooperation with the Voltaix company. The enrichment of this silane gas was as follows: $^{28}$Si (∼99.924\%), $^{29}$Si (∼0.073\%), and $^{30}$Si (∼0.003\%) \cite{Itoh14}. In our experiments we have used silane produced from $^{28}$SiF$_4$ according to \cite{Bulanov00}. The details of its synthesis and characterization are in short described below.

 \section{Experimental methods}
 
  \subsection{Fabrication of the $^{28}$SiH$_4$ precursor}
 
%We choose the Type 3 method    ($^{iso}$SiF$_4$ $\rightarrow$ $^{iso}$ SiH$_4¤$ multistep Fluoride-Hybride fabrication \cite{Bulanov00}) to produce high isotopic purified   $^{iso}$ SiH$_4¤$ in large enough quantity compatible with our pre-industrial CVD 300 mm equipment.
%In fact $^{iso}$SiF$_4$ can be purified up to very high grade using 
%centrifugation. The world record is $^{iso}$SiF$_4$ at $^{28}$Si= 99.9999664(1) \% \cite{Sennikov12}.

Gas centrifugation is the only effective method of isotope separation with a high enrichment level. Other methods, such as magnetic mass separation, ion exchange and laser technology, are more expensive. These methods do not allow a high isotopic enrichment to be reached. There are some limitations for gases to be used in gas centrifuges. One of the most important limitations is the molecular mass of the gas. Silane is too light for this purpose. This is why silicon tetrafluoride SiF$_4$ is used as a process gas for the centrifugal separation. Two other reasons are as follows:  only one isotope exists for fluorine and this substance has sufficiently high vapour pressure at room temperature. Synthesis and enrichment of SiF$_4$ took place in Russia at SC "PA Electrochemical Plant" (ECP), Zelenogorsk, Krasnoyarsk region. The natural abundance silicon ($^{28}$Si $\simeq$ 92.23 \%) of electronic grade quality was manufactured and supplied by Wacker Polysilicon Europe, Wacker Chemie AG, Germany. SC “PA ECP” produced high-purity fluorine (F$_2$) by itself. A special installation with a reactor was constructed and used for the synthesis of the initial 
SiF$_4$ and for silicon isotope separation in centrifugal cascades. Other details of enrichment procedure and purification of 
SiF$_4$ are given in \cite{Abrosimov17}.

Then $^{28}$SiF$_4$  was converted into silane, $^{28}$SiH$_4$. First, the synthesis of silane was carried out by the reaction of high-purity silicon tetrafluoride with calcium hydride:

$$ ^{28}SiF_4 + 2CaH_{2}\rightarrow  ^{28}SiH_{4} + 2 CaF_{2}$$			

A mixture of isotopically enriched silicon tetrafluoride with hydrogen of special grade B purity was passed through a layer of mechanically dispersed calcium hydride. The reactor was made of high-purity Si-free stainless steel to prevent the diffusion of boron compounds and natural Si into the highly enriched gas. The content of hydrocarbon impurities in silane after synthesis was at the level of 10$^{-5}$ mol/mol. The concentration, around 10$^{-3}$ mol/mol, of polysilanes and disiloxanesis was the largest component according to  gas-chromatographic-mass-spectrometric (GC-MS). Calcium hydride seems to be the main source of impurities, in particular carbon.
The produced $^{28}$SiH$_4$ was freed from impurities by  cryofiltration and periodic low-temperature rectification.

 \subsection{Growth of the $^{28}$Si epilayer}
 
 	A 300mm Epsilon 3200 single wafer tool from ASM America was used for the epitaxial growth of natural and isotopically pure Si layers. Software and hardware modifications were implemented on the tool to be able to (i) switch from natural to isotopically pure SiH$_4$ and (ii) deliver precise amounts of silane (whatever its nature) in the epitaxy chamber thanks to a Mass-Flow Controller (MFC). Given the cost of $^{28}$SiH$_4$, a dedicated gas box was installed in the basement of our cleanroom as close as possible to the tool to minimize the length of the gas line 
 	($\simeq$  10 m against $\simeq$ 70 m).
	As we will be targeting in the coming years the deposition of layers several nm to several tens of nm thick on bulk Si or Silicon-On-Insulator substrates, we have decided to investigate the properties of layers grown at 650$^{\circ}$C, 20 Torr with a F($^{28}$SiH$_4$)/F(H$_2$) mass-flow ratio of 0.012. The Si growth rate is then typically around 10 nm min.$^{-1}$ \cite{Hartmann12}, which should be handy for thin layers and the temperature low enough to enable epitaxy on thinned-down SOI substrates, with a starting Si layer thickness as low as 3 nm \cite{Jahan05}.

 \section{Results}
 \subsection{Characterization of the $^{28}$SiH$_4$ precursor.}
 
 The isotope concentrations in silane in the transport vessel were measured by laser source mass spectroscopy (LIMS) and are as follows: $^{28}$SiH$_4$= 99.99748 ($\pm$ 0.00026) at.\%, $^{29}$SiH$_4$= 0.00250 ($\pm$ 0.00026)  at.\% , $^{30}$SiH$_4$= 0.00002 ($\pm$ 0.00001) at.\%.

 Chemical analysis of unintentional species was performed  by gas chromatography  and chromato-mass-spectroscopy from the transport vessel. High-purity $^{28}$SiH$_4$ was prepared with an overall hydrocarbon content less than 2 $\times$ 10$^{-7}$ mol/mol, less than   4 $\times$ 10$^{-8}$ mol/mol of disiloxane and around 10$^{-7}$ mol/mol of higher order silanes (see table \ref{table1}).
 \begin{table}
 
 \begin{center}
 \begin{tabular}{| p{3cm} | p{3cm}| }
  \hline
  species & concentration \\
   & ($\mu$mol/mol)  \\
  \hline
  CH$_4$ & $\le$ 0.05  \\
  C$_2$H$_6$ & $\le$ 0.02 \\
  C$_2$H$_4$ & $\le$ 0.02 \\
  C$_3$H$_6$ & $\le$ 0.02 \\
  C$_3$H$_8$ & $\le$ 0.01 \\
  i-C$_4$H$_{10}$ & $\le$ 0.02 \\
  n-C$_4$H$_{10}$ & $\le$ 0.02 \\
  More than 52 other species & $\le$ 80.35 $\pm$ 20.135\\
  \hline
\end{tabular}
\caption{hydrocarbon contents and others species (gas chromatography) in the silane (from the transport vessel)}
\label{table1}
\end{center}
\end{table}

 \subsection{Isotopical, Chemical, morphological and physical characterization of the $^{28}$Si epilayer.}
 
We have first  profiled with secondary ion mass spectrometry (SIMS) the various Si isotopes in a ~ 60 nm thick Si layer grown at 650$^{\circ}$C, 20 Torr with $^{28}$SiH$_4$ on a natural abundance Si(001) wafer (after a 1100$^{\circ}$C, 2 min. H$_2$ bake in order to get rid of native oxide).
 
\begin{figure}[t]
\begin{center}

\includegraphics[width=8.5cm]{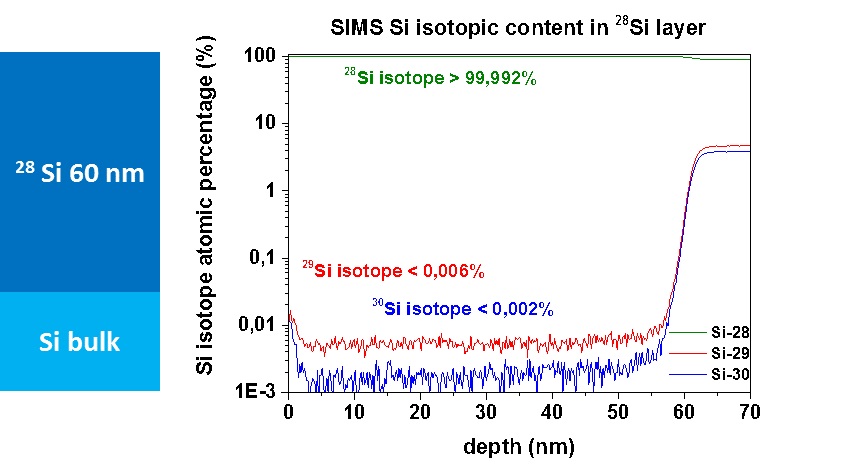}
\caption{(color online). Secondary ion mass spectrometry  measurements. Left panel: targeted layer composition. Right panel: (Top)  SIMS  depth profiling of the Si isotopes in the resulting layer in Grenoble. The concentration of the $^{29}$Si isotope which contains a nuclear spin $1 \over 2$ is less than 0.006 \% in the top layer and the concentration of the nuclear spin-free $^{30}$Si isotope lower than 0.002 \%.   SIMS depth profiling of the Si isotopes analysis in  Nizhny-Novgorod (not shown) gives: $^{28}$Si=99.99337 ($\pm$ 0.001) at.\%,  $^{29}$Si=0.00524  ($\pm$ 0.0009) at.\% , $^{30}$Si=0.00139  ($\pm$ 0.0005) at.\% for the epilayer, in very good agreement with experiment in Grenoble. Both were performed on ION TOF TOF-SIMS 5 instruments. There is otherwise  a small isotopic dilution when going  from the natural abundance Si(001) wafer towards the $^{28}$Si epilayer.}
\label{fig1}
\end{center}
\end{figure}

 SIMS measurements of silicon isotope concentrations in Grenoble and Nizhny Novgorod (IPM RAS) were close to each other (see figure \ref{fig1}). 
 
 %and indicate on small isotopic dilution as going from starting silane ($^{28}$SiH$_4$= 99.99748 ($\pm$ 0.00026) at.\%)  to epilayer ($^{28}$Si=99.99337 ($\pm$ 0.001) at.\%, see figure \ref{fig1}). Though, due to this dilution twofold  increase of the concentration of $^{29}$Si isotope takes place:
 The $^{28}$Si isotopic concentration in the epilayer from SIMS, 99.99337 ($\pm$ 0.00026) at.\% was slightly lower than the LIMS  in the 
 $^{28}$SiH$_4$ bottle, i.e. 99.99748 ($\pm$ 0.00026) at.\%. This was associated with a two fold increase of the harmful 
  $^{29}$Si isotopic concentration from 0.00250 ($\pm$ 0.00026)  at.\%  up to  0.00524 ($\pm$ 0.0009) at.\% (from LIMS to SIMS).
  % The previous deposition of natural silicon epilayers in the chamber is probable responsible for this dilution. Very small increase of $^{29}$Si and $^{30}$Si concentration inside 10 nm layer close to substrate is due to diffusion of natural silicon from this substrate.  

Metrology techniques were not the same, however, which might partly explain those differences. The epitaxy chamber quartz wall, the SiC-covered susceptor plate (on which wafers lay during growth) and other inner components are cleaned between depositions thanks to high temperature, high pressure etches with gaseous HCl. A memory effect stemming from the epitaxy beforehand of a large number of natural abundance Si layers cannot be fully excluded, however. Finally, although the fraction of the gas panel common to natural and purified silane was purged between switches, some unwanted molecules might have been still present in the  MFC and in the injection line during the deposition of that $^{28}$Si layer. The increase of the $^{29}$Si and $^{30}$Si isotopic concentrations in the 10 nm close to the substrate is likely due to the diffusion of natural silicon from the latter towards the epilayer.

In order to gain access to the Si growth rate with those process conditions and assess the structural quality of Si layers grown with  $^{28}$SiH$_4$, we have proceeded as follows. We have started from natural abundance Si(001) substrates and grown at 650$^{\circ}$C, 20 Torr (SiGe 30\% 19 nm / Si cap 30 nm) stacks. As before, a 1100$^{\circ}$C, 2 min. H$_2$ bake was used beforehand to get rid of native oxide. A SiH$_2$Cl$_4$ + GeH$_4$ chemistry was used for the growth of the SiGe 30\% marker layers, while natural abundance or purified silane was used for the growth of the Si caps.

\begin{figure}[t]
\begin{center}

\includegraphics[width=8.5cm]{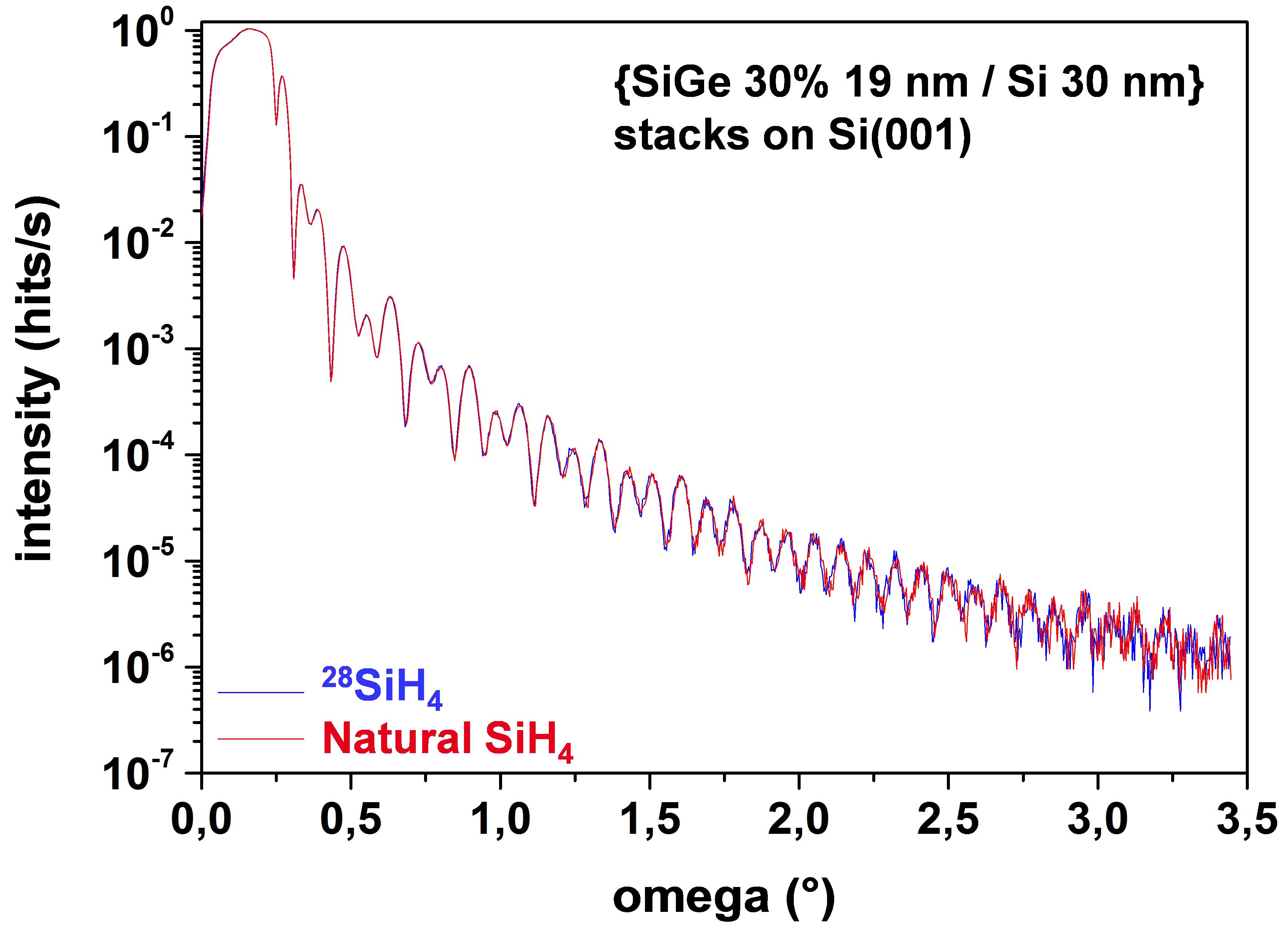}
\caption{(color online). X-Ray Reflectivity profiles associated with (SiGe 30\% 19 nm / Si 30 nm) stacks grown at 650$^{\circ}$C, 20 Torr on Si(001). The 30 nm thick Si caps were grown using natural or purified silane.  }
\label{fig2}
\end{center}
\end{figure}
We have first  used X-Ray Reflectivity (XRR) to (i) gain access to the individual layer thickness of those stacks (and thus determine the Si growth rate) and (ii) qualitatively see if the Si caps were smooth. XRR profiles are provided in Figure \ref{fig2}. 
	The presence of well-defined thickness fringes up to high incidence angles is a clear sign that the surface of the Si cap was smooth and the interfaces (between SiGe 30\% and Si) abrupt irrespectively of the silane source used. The presence of two sets of thickness fringes is due to constructive and destructive interferences of the X-rays reflected at the {Si cap / SiGe 30\%} and the {SiGe 30\% / Si substrate} interfaces as omega, the angle of incidence, changes. Their angular period is inversely proportional to (i) the Si cap thickness and (ii) the {SiGe 30\% / Si cap} thickness. We extracted from the Si cap thickness the following Si growth rate: 10.1 nm min.$^{-1}$. The perfect superposition of the XRR profiles is a clear sign that growth proceeds exactly the same way with both sources of silane.

	We have completed the characterization of those stacks with tapping-mode Atomic Force Microscopy (AFM) and surface haze measurements (in Bruckers FastScan and KLA Tencor SP2 metrology tools, respectively).

\begin{figure}[t]
\begin{center}

\includegraphics[width=8.5cm]{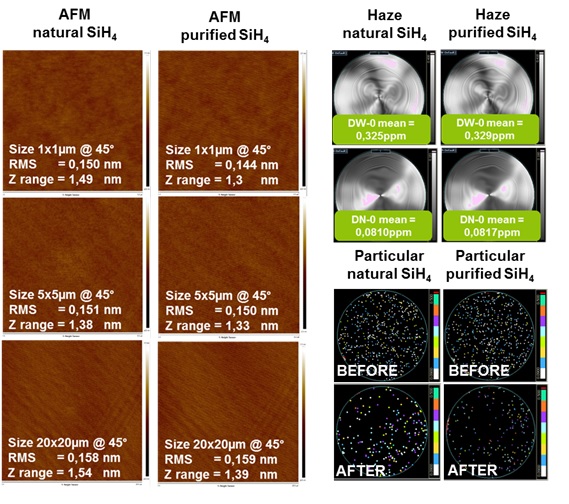}
\caption{(color online). Left panel: Tapping-mode Atomic Force Microscopy images of the surface of the 30 nm thick Si caps grown at 650$^{\circ}$C, 20 Torr with natural or purified silane. Smooth, featureless surfaces were obtained in both cases whatever the scan size. Scan direction was along $<$100$>$. Top right panel: DW and DN Haze maps. Bottom right panel: particular 
contamination maps of the wafer surface before and after epitaxy for the two kinds of Si caps, with minimum particule size at 90nm }
\label{fig3}
\end{center}
\end{figure}

Figure \ref{fig3}, left panel shows various scan size (1 $\mu$m x 1 $\mu$m, 5 $\mu$m x 5 $\mu$m and 20 $\mu$m x 20 $\mu$m) AFM images of the surface of Si caps grown using either natural silane (left column) or purified silane (right column). Smooth, featureless surfaces were obtained whatever the scan size and the source of silane, with a Root Mean Square (RMS) roughness and a Z range (Zmax.--Zmin.) equal on average to 0.153 nm and 1.47 nm for natural silane and 0.151 nm and 1.34 nm for purified silane.
	Figure \ref{fig3}, top right shows Dark Wide (DW) and Dark Narrow (DN) haze maps associated with diffuse light scattering (by surface roughness) by our 300mm wafers in a SP2 tool \cite{Hartmann11}. The DW and DN haze values are very low and once again strikingly similar (DW haze = 0.325 parts per million (ppm) and DN haze = 0.0810 ppm for natural silane, to be compared with 0.329 ppm and 0.0817 ppm for purified silane). We thus confirm at the wafer scale the conclusions reached with XRR and AFM, i.e. that the Si cap surfaces are smooth at the atomic scale and featureless whatever the source of silicon.
	Finally, we have also had a look, in our SP2 tool, at the particules (which might be large size epitaxy defects coming from imperfect gas sources) detected after the epitaxy of those stacks. Maps at the wafer scale are provided in Figure \ref{fig3}, bottom right. The number of counts and defects are quite similar whatever the source of silane, which is very reassuring.

We have otherwise assessed whether or not we were faced with metallic impurities after the growth of numerous $^{28}$Si layers on various types of substrates. It might indeed have happened that, because of the numerous very specific process steps used to obtain $^{28}$SiH$_4$, the gas was contaminated. To that end, we have loaded,
after the growth of numerous $^{28}$SiH$_4$ layers,  bulk Si wafers inside the epitaxy chamber, left them in it for several minutes at 900$^{\circ}$C, 760 Torr and unloaded them. We have then used Total X-Ray Fluorescence (TXRF) spectroscopy to detect a metallic contamination on the front side and the back side of those wafers. We were in every case below the TXRF detection threshold (typically around 10$^{10}$ atoms/cm$^2$) for the various metals tracked, i.e.  Al, Ca, Co, Cr, Cu, Fe, K, Mg, Mn, Na, Ni and Ti. The number of added particules on the front side (120 nm detection threshold in the SP2 tool) after this loading/unloading was barely 3.

From all those experiments, we can safely state that the purified silane source we used yields the same epitaxial quality layers than microelectronics-grade silane, without any deleterious metallic and/or particular contamination afterwards. The isotopic purity in epilayers is very high, which should definitely be an advantage in quantum electronics. In the next section, we will quantify, in specially tailored stacks, the impact of thermal budget on the self-diffusion of Si.

\begin{figure}[t]
\begin{center}

\includegraphics[width=8.5cm]{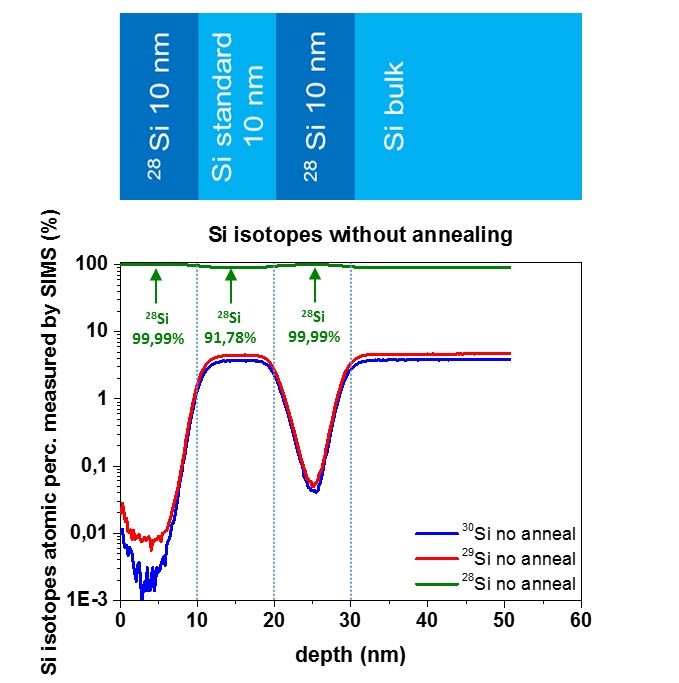}
\caption{(color online). Si isotopes  atomic percentages from SIMS in a {$^{28}$Si 10 nm / natural or “standard” Si 10 nm / $^{28}$Si 10 nm} stack grown at 650$^{\circ}$C, 20 Torr with $^{28}$SiH$_{4}$ and natural Si$_{2}$H$_{6}$ on a bulk Si(001) substrate.}
\label{fig4}
\end{center}
\end{figure}

\section{Si self-diffusion in natural Si / $^{28}$Si heterostructures}

In order to confirm that Si self-diffusion was the same in our CVD-grown layers and in the literature, we have deposited at 650$^{\circ}$C, 20 Torr on 300mm Si(001) wafers the following stacks (from bottom to top) 10 nm of $^{28}$Si / 10 nm of natural Si / 10 nm of $^{28}$Si. It is not practical to switch from natural silane to purified silane in the same growth run. Using dichlorosilane (SiH$_{2}$Cl$_{2}$) is otherwise not an option as the Si growth rate at 650$^{\circ}$C is prohibitively low. We have thus  taken advantage of the fact that we also have natural disilane (i.e. Si$_{2}$H$_{6}$) connected to our tool to grow the natural Si layer, which is sandwiched between the two $^{28}$Si layers. The natural disilane mass-flow used, the lowest one that could reproducibly be delivered in our tool, was 11 times less than that of purified silane. The 650 $^{\circ}$ C, 20 Torr growth rate from disilane was nevertheless twice higher (i.e. 20 instead of 10 nm min.$^{-1}$).The SIMS depth profile of the $^{28}$Si, $^{29}$Si and $^{30}$Si isotopes in the as-grown stack can be found in Figure \ref{fig4}. Several things are obvious: 

(i) The $^{29}$Si and $^{30}$Si concentrations are, as expected, the same in the 10 nm thick natural Si layer than in the Si substrate; 

(ii) Most likely because of intermixing in the SIMS crater during profiling, the $^{29}$Si and $^{30}$Si concentrations are nearly one decade lower in the $^{28}$Si cap than in the bottom $^{28}$Si layer. The latter layer was indeed profiled after the natural Si spacer. 

(iii) The chemical width of the interface is several nm only, i.e. the same kind of width found during the Atom Probe Tomography of Si/SiGeC superlattices grown also at 650$^{\circ}$C \cite{Estivil15}. We indeed have steady $^{29}$Si and $^{30}$Si concentrations in the majority of the top $^{28}$Si layer and the natural Si layer just below. This is important for device processing, as it means that a “natural” SOI substrate could behave almost as a $^{28}$SOI substrate provided that a $^{28}$Si epilayer 10 nm thick or more is deposited on top.

\begin{figure}[!h]
\begin{center}

\includegraphics[width=8.5cm]{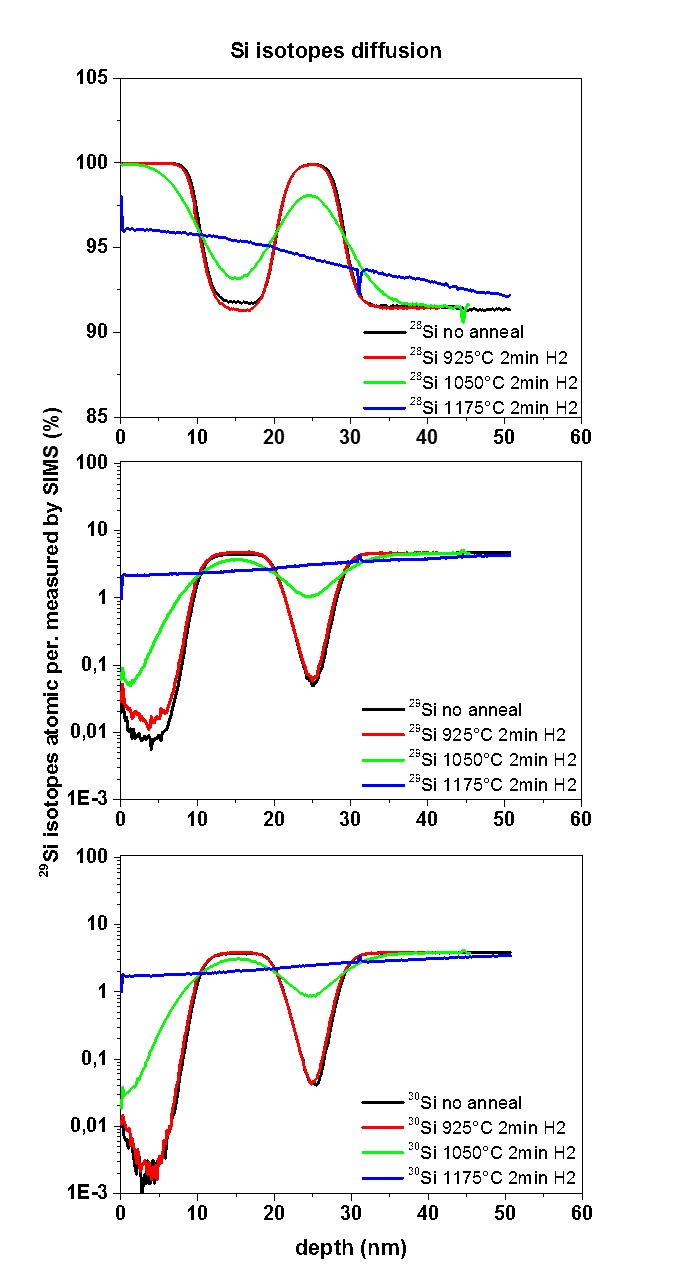}
\caption{(color online). Si isotopes’ atomic percentages from SIMS in a {$^{28}$Si 10 nm / natural or “standard” Si 10 nm / $^{28}$Si 10 nm} stack (i) grown at 650$^{\circ}$C, 20 Torr on a bulk Si(001) substrate (no anneal) or (ii) annealed for 2 min. under 
H$_{2}$ at 925$^{\circ}$C, 1050$^{\circ}$C or 1175$^{\circ}$C. It should be highlighted that the y-scale, linear for $^{28}$Si, is logarithmic for $^{29}$Si and $^{30}$Si isotopes.}
\label{fig5}
\end{center}
\end{figure}

Having seen that, we have submitted the same stacks to 2 min. H$_{2}$ anneals at various temperatures (925$^{\circ}$C, 1050$^{\circ}$C and 1175$^{\circ}$C) in the epitaxy chamber to emulate the thermal budget stacks made of natural Si and $^{28}$Si would be submitted to during device fabrication. The ramping-up and ramping-down rate (from 550$^{\circ}$C up to the annealing temperature), 2.5$^{\circ}$C/s, was low enough to minimize temperature overshoot. SIMS depth profiles of the $^{28}$Si, $^{29}$Si and $^{30}$Si isotopes can be found in figure \ref{fig5}. The $^{29}$Si and $^{30}$Si depth profiles are almost the same for the as-grown sample and for the sample annealed for 2 min. at 925$^{\circ}$C. Moving over to 1050$^{\circ}$C has a huge impact on isotopic profiles, which are flattened with definitely broader interfaces from the chemical point of view. Annealing such a stack at 1175$^{\circ}$C for 2 min. completely suppress the fluctuations because of intermixing. The only differences with bulk Si are then, as expected, the isotopic concentrations which are different from the natural ones in the first 50 nm or so of Si (the $^{28}$Si isotope is more abundant and the $^{29}$Si and $^{30}$Si isotopes less abundant, then, which is logical given that 20 nm of $^{28}$Si was deposited). 
 
 \begin{figure}[t]
\begin{center}

\includegraphics[width=8.5cm]{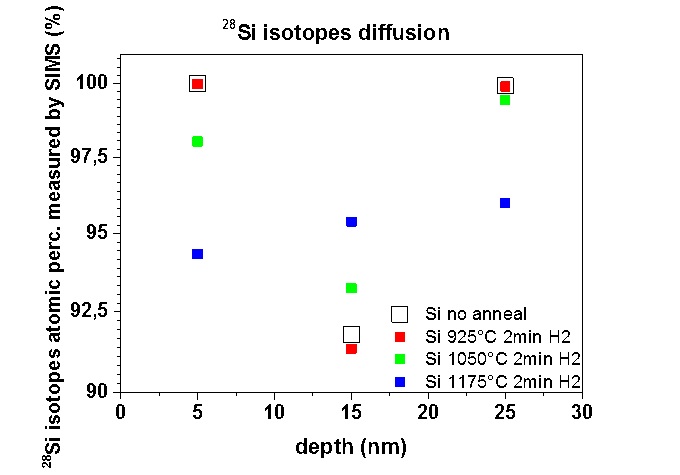}
\caption{(color online). $^{28}$Si isotopic concentration in the middle of the two $^{28}$Si layers and in the natural Si spacer in-between for the as-grown sample or the stacks annealed for 2 min. under H$_{2}$ at 925$^{\circ}$C, 1050$^{\circ}$C or 1175$^{\circ}$C.}
\label{fig6}
\end{center}
\end{figure}

 We have extracted what could be called "steady state" $^{28}$Si isotopic concentrations in the middle of each 10 nm thick layer (i.e. at 5 nm, 15 nm and 25 nm SIMS depths) for the various samples investigated. Data are plotted in Figure \ref{fig6}. While $^{28}$Si concentrations are barely lower in the two $^{28}$Si layers after a 2 min. H$_{2}$ anneal at 925$^{\circ}$C than in the as-grown layers, it is obvious that concentrations go back to the natural abundance for higher thermal budgets.

\begin{figure}[t]
\begin{center}

\includegraphics[width=8.5cm]{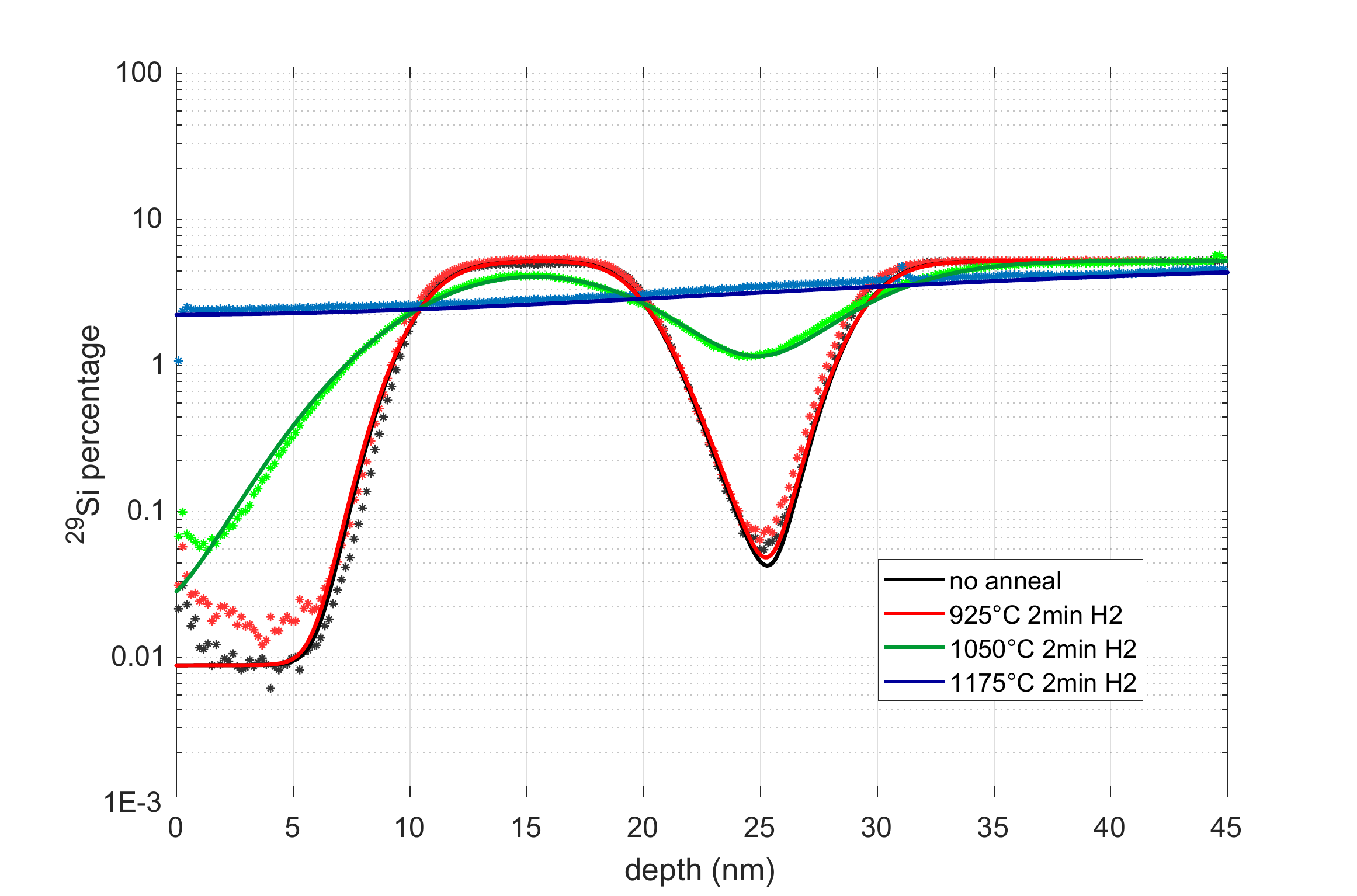}
\caption{(color online). SIMS data (symbols) and simulation results (lines) of Si self-diffusion. The depth resolution function based on MRI-model applied to simulated profiles uses an atomic mixing w = 0.95 nm, surface roughness $\sigma$ = 1.3 nm and information depth $\lambda$ = 0.39 nm. }
\label{fig7}
\end{center}
\end{figure}

A 1D finite difference time domain (FDTD) model for the calculation of Si self-diffusion was used to simulate the SIMS profiles shown in figure \ref{fig5}. We assume infinitely sharp interfaces between the $^{28}$Si and $^{nat}$Si layers of the stacks. 
Temperature ramp-up and down of the annealing are included in the simulations. For the self-diffusion coefficient, a single Arrhenius law:
$  D_{Si}^{SD} = 423 \times exp(- {4.73 eV \over k_{B} T}) cm^{2}s^{-1}$ 
is assumed, following the results of ref. \cite{Suedkamp16}.
 A standard MRI (ion Mixing surface Roughness Information depth, first developed in
 ref. \cite{Hofmann94}) depth resolution function is then convoluted to the simulation output to reproduce the finite depth resolution of SIMS data. 
 The MRI parameters are set with a single fitting procedure to the whole data set (see figure caption figure \ref{fig7}). As can be seen on figure \ref{fig7}, the measured $^{29}$Si SIMS profiles are in good/excellent agreement with our simulations based on previously reported Si self-diffusion coefficients. 

 \section{Conclusion}
 
We reported  the growth of high structural quality  $^{28}$Si crystalline layers with enrichment $\ge$ 99,992 \% and with a very low level of other contaminants.
The isotopically purified epilayers are grown on 300mm substrates which are the  standard wafers for CMOS foundries. The epilayers have been characterized in depth by several complementary techniques in different places. The level of $^{28}$Si enrichment is larger than in previous reports in the context of silicon qubits and constitutes a record for silicon films of large area. A study of the isotopic concentration profile as function of annealing temperature  is presented which permits to identify the allowed thermal budget range for the  subsequent   qubit fabrication.

 \section{Acknowledgements} 
 This work was supported by  the EU H2020 program (under the MOSQUITO  project No 688539), by the Russian  Federal Agency of Scientific Organizations ( project No AAAA-A17-117030910056-1) and by the Russian Foundation for Basic Research (Project No 18-03-00235).

\end{document}